# STRONGLY ANISOTROPIC S-WAVE GAPS IN EXOTIC SUPERCONDUCTORS

B. H. Brandow

Center for Nonlinear Studies

Los Alamos National Laboratory

Los Alamos, NM 87545

The exotic superconductors, defined as those which follow the phenomenological trend of Uemura ($T_c$ approximately $\propto \lambda_L^{-2}$), presently constitute the most broad and general class of superconductors which can reasonably be considered "similar to the high-$T_c$ cuprates". It is therefore of much interest to determine the forms of their pairing gap functions. We examine evidence for the gap forms in non-cuprate exotics, and demonstrate some general features. The cubic materials often have highly anisotropic gaps. The planar materials tend to have even stronger gap anisotropy, to the extent that they often have gap nodes, but nevertheless they usually have an s-wave-like gap symmetry. There is good evidence for the latter even in the controversial cases of planar organics and nickel borocarbides. Exceptions to these generalizations are also pointed out and discussed.

§1. INTRODUCTION

There are now many unusual superconductors which are recognized to share some of the novel features of the high-$T_c$ cuprate superconductors. The possibility of further similarity to the cuprates, in particular the possibility of pairing by exchange of a virtual spin fluctuation or paramagnon, has motivated suggestions of d-wave gap forms in some of these other superconductors, in cases where there is evidence for gap nodes. This argument has been applied especially to the planar organics and sometimes also to the nickel borocarbides, in addition to the well known cases of heavy-fermion materials. This line of reasoning calls for a careful examination of the evidence for the gap forms in these and other unusual superconductors.

In this paper we focus on the exotic superconductors of Uemura and co-workers.[1] These exotics include a number of the unusual superconductors of earlier (pre-cuprate) interest, as well as the cuprates themselves and others discovered after the cuprates, and they include most of the materials previously suggested as being similar to the cuprates.[2] As discussed below, they now constitute the most broad and general class of superconductors that can reasonably be considered similar to the cuprates. (The variety of crystal structures and material chemistries in this class of materials is remarkably extensive.[2]) It is therefore very reasonable to focus on these materials. We now examine many of the non-cuprate exotic materials and material families, those for which there is reasonable gap-form evidence, in order to establish two general conclusions about their superconducting phenomenology: (1) In the cubic and nearly-cubic materials, which are conventionally expected to have nearly isotropic gaps, there is often evidence of *very strong* gap anisotropy. Sometimes there even appear to be gap nodes, although the gap symmetry remains s-wave-like. (2) In the planar (quasi-two-dimensional) materials there are often gap nodes, but nevertheless they typically still have an s-wave-like gap symmetry. [Here s-wave-like means having the full point-group symmetry of the crystal structure, a condition which does allow the possibility of gap nodes (Mahan 1989, Fedro and Koelling 1993).] Altogether, therefore, there is much more similarity throughout the cubic and the planar materials than is commonly recognized. The gap nodes are seen to arise merely because of strong gap anisotropy, and not because of a change of the overall gap symmetry. This is the general trend. There are some exceptions, however, which we also point out and discuss. (Logically we should also examine quasi-one-dimensional examples, but we have not done so.) This survey can be compared to an earlier gap-form study of many of the same materials (Annett 1999). Our orientation is somewhat different and many of the individual conclusions differ, based on a thorough examination of the available data.

The background for this investigation is an evolution from the many studies that have recognized phenomenological similarities among various unusual superconductors, often but not always comparing these materials to the high-$T_c$ cuprates. A number of these studies predated the discovery of the cuprates. [Noteworthy examples of such studies are listed in Ref. 5 of Brandow (1998).] As just mentioned, the cuprates as well as most of the previously suggested "similar to cuprates" materials are among the exotic superconductors. These exotic materials were so-named because they were found to follow a quite unexpected trend, namely $T_c$ approximately proportional to $\lambda_L^{-2}$, where $\lambda_L$ is the London penetration depth (for $T \rightarrow 0$). On a log-log plot this trend is essentially orthogonal to the trend of the elemental superconductors, which demonstrates that these unusual superconductors are indeed quite special. They are clearly different from the elemental superconductors, and they are obviously sharing something in common. It is therefore very reasonable to infer that some new or unconventional pairing mechanism must be active throughout these materials. (Nevertheless there is evidence that the conventional phonon mechanism is also active in these materials, at least in most cases; see comments below. The heavy-fermion materials are apparently exceptions in this regard, and also $Sr_2RuO_4$; see below.) We consider this Uemura trend to be the most convincing single type of evidence for a shared new pairing mechanism. In a field with so many conflicting claims, the Uemura relation stands out as a trustworthy basis for further study. The Uemura trend is discussed further in Appendix A.

This strong indication for a shared new pairing mechanism has led us to carefully examine other phenomenological similarities among these exotic materials, exploring and extending the more fragmented observations of many previous works (Brandow 1998).[3] Consistent with and perhaps exceeding the previous expectations, a remarkably broad commonality among these exotic materials was found, a general similarity which involves many aspects. (The characteristic features involve superconducting and normal-state electronic properties, and also crystal-chemistry features.) This finding strongly supports and reinforces the idea of a shared unconventional pairing mechanism. At the same time this provides substance and clarity for the previously vague notion of being "similar to the cuprates". We must emphasize that in spite of the cuprates having some apparently unique features (linear resistivity, stripes, and pseudogap phenomena), the cuprates clearly do share most of the characteristic features of the other exotic superconductors. The cuprates are not as unique as has sometimes been claimed.

Consistent with this broad commonality, there is actually much evidence for a highly anisotropic (and node-containing) *s-wave-like gap form in the cuprates*, for both the hole-doped and electron-doped cases. We have recently reviewed this evidence, and have also demonstrated that some of the apparently strongest d-wave evidence, from phase-sensitive Josephson tunnelling experiments, is not inconsistent with an anisotropic s-wave gap form (Brandow 2002; see also Klemm 1999, Bille *et al*. 2001, Zhao 2001a). We are thus arguing that an s-wave-like gap form, usually with strong anisotropy and often with nodes, is a feature of most of the exotic superconductors – another one of the many characteristic features – and that this quite possibly and even probably applies also to the cuprates. Further evidence for the s-wave form in cuprates is discussed here in the concluding section.

A likely possibility, we believe, is that the total set of exotic materials, including those yet to be discovered or confirmed, may be simply the set of strong type-II superconductors. The exotics generally have quite short coherence lengths $\xi$ and long penetration depths $\lambda_L$, and thus they have large values (greater than or about 10) for the Ginzburg-Landau parameter $\kappa = \lambda_L/\xi$. Another prominent characteristic of these materials is that they are all chemical compounds, not single elements, and they usually contain strongly electronegative elements that by themselves are non-metallic or semi-metallic. Strongly electropositive elements also sometimes enter. It is therefore typical to find strong differences among the electronegativities of the constituent elements. (Smaller electronegativity differences are sometimes found in heavy-fermion superconductors, where all of the constituent elements may be metals, but other energy scales are also smaller in these materials.) These chemical features are counter-intuitive for superconductivity, and this has led us to characterize these materials as "strange-formula superconductors" (Brandow 1998). We mention these observations because there are many other superconductors with these strong-type-II and strange-formula features, some of these materials being long known[4] and others only recently discovered, and these other materials are all good candidates for exotic status. In this paper we focus mainly on the "officially exotic" materials which have been confirmed to follow the $T_c \propto \lambda_L^{-2}$ trend (confirmed by μSR, muon spin relaxation data), but we also include some of the other "obvious candidate" materials.

The example of the cuprates will serve to illustrate the present concept of an anisotropic s-wave gap form:

$$\Delta(\theta) \propto (1 + r\cos 4\theta) \quad . \tag{1}$$

In the cuprates the angle $\theta$ parametrizes the position on the "Fermi circle" as viewed from the enclosed $(\pi,\pi)$ point, the angle being measured from a planar Cu-O bond direction. This gap is maximum along the Cu-O bond direction, in agreement with the commonly assumed $x^2-y^2$ d-wave gap form. Assuming $r > 1$, this $\Delta$ has a pair of nodes near each diagonal or (q,q) direction, for a total of eight gap nodes, and there is also a subsidiary or secondary maximum of negative sign at each diagonal point on the Fermi circle. This form was suggested long ago by Mahan (1989). This particular (cuprate case) gap form can be described as "s+g", with $s$ being the isotropic component and $g$ labelling the $l = 4$ component ($\cos 4\theta$ component). It can also be called an "eight node" form. This might also be called an "extended s" form, but we avoid that term because this is often attached to a precise form which is too restrictive for the present context. "Nodal s-wave" is appropriate when nodes are present, but for the general case (with or without nodes) we shall refer to this simply as an anisotropic-s-wave gap form.

Although the particular gap form (1) may apply to few other exotic materials (e.g. the planar organics; §3), this form serves to illustrate how gap nodes can occur within an overall s-wave-like gap symmetry (tetragonal $A_{1g}$ symmetry for the cuprates), where the gap has the full point-group symmetry of the lattice. In the cuprate context we have emphasized that there is an unfortunate but widespread tendency to claim gap-node evidence as being evidence for a d-wave gap form (Brandow 2002). This is wrong, as (1) clearly shows. This same false logic is also very frequently applied to other planar exotics. Another common mistake is to assume that the "old" superconductors (such as A-15's, $CeRu_2$, $NbSe_2$) must have nearly isotropic gaps, either because the materials are cubic or nearly cubic, or because of the presumption that "old superconductors must be conventional". These assumptions are refuted by the evidence presented here.

Several experimental consequences follow immediately from this type of gap form. The first is the presence of a subsidiary or "within the gap" peak in the one-electron tunnelling state density, if the anisotropy [*e.g.* the *r* parameter in (1)] is strong enough to produce negative-sign regions in the gap (Mahan 1989, Fedro and Koelling 1993). Such a peak has been clearly observed in tunnelling for several cuprates (Brandow 2002 and references therein, Zhao 2001a,b), and there is also less distinct evidence for this in much of the cuprate tunnelling data. To our knowledge, however, no such peak has been clearly seen in other materials. (A variety of reasons could make it difficult to observe such peaks; they are often not clearly seen even in the cuprates, where the large gap magnitude is especially favorable for this.) Another consequence is that this type of gap, even with strong anisotropy, is still symmetry-compatible with the conventional phonon mechanism. This means that the phonons can and often apparently do also contribute substantially to the pairing. (The evidence is mainly from tunnelling, in structure beyond the gap.) It can therefore be difficult to determine whether something *other than* the phonon mechanism is contributing to the pairing, if one relies only on the conventional phonon-based methodology for analyzing a superconductor. Indeed, a majority of the materials discussed in this paper have been and still are widely believed to be conventional superconductors. The contrary evidence is not sufficiently well known.

For the purpose of this paper the most important consequence of this type of gap form is that gap nodes can be removed in a very simple and straightforward manner, by merely reducing the gap anisotropy. This can be brought about either by some nonmagnetic "dirt" or disorder in the sample, which tends to angle-average the gap function, or by altering some of the basic electronic parameters.[5] Elimination of gap nodes is therefore much more straightforward here than for the case of a d-like (or other $l \neq 0$) gap form. [In the d-wave scenario, an avoidance or removal of gap nodes is typically attributed to a violation of time-reversal invariance. But in the case of the cuprate YBCO ($YBa_2Cu_3O_{7-\delta}$) there is evidence against time-reversal violation (Mathai *et al*. 1995). This method of node removal is likewise more straightforward than the suggestion of a d-wave to s-wave transition, which is discussed in the concluding section.] The elimination of gap nodes by either of these means (dirt/disorder or electronic alteration) is therefore strong evidence for an s-like gap symmetry. Granted that such evidence is not a definitive proof, in view of the alternative possibilities just mentioned, its simple and straightforward basis does require that this evidence be taken very seriously. This argument has been used previously for cuprate materials (Hotta 1993 and Norman 1994 for hole-doped cuprates; Brandow 2002 for electron-doped cuprates), and we present a number of further examples here. We have now described the most obvious consequences of the present type of gap form. More subtle consequences are discussed in Brandow (2002).

Although we find that s-wave-like gap symmetry is typical for the exotic superconductors, there are some exceptions. These are the uranium-based heavy-fermion superconductors, for some of which the highly unconventional feature of multiple superconducting phases is well known, and also $Sr_2RuO_4$ for which the Cooper pairs apparently have triplet spin (S = 1). [See for example Sigrist *et al*. (1999), Mackenzie and Maeno (2000), Maeno *et al*. (2001).] We have pointed out elsewhere (§5.2 of Brandow 2000) that from a valence-fluctuation perspective the uranium materials differ from most of the other exotic superconductors. The uranium ions in these materials fluctuate mainly between the single-ion configurations $f^1$, $f^2$, and $f^3$, *all* of which have non-vanishing local moments. This makes the uranium-ion case inherently more complicated than, say, the related cerium materials. (In the valence-fluctuation description of most of the other exotics, each active ion is forced into a singlet state by the fact that at least one of its strongly-admixed ionic configurations is a singlet.) It is therefore quite reasonable that the pair wavefunctions of these uranium-ion materials can exhibit degrees of freedom not found in most of the other exotic materials. This argument also applies to $Sr_2RuO_4$, where the nominal Ru configuration is $d(t_{2g})^4$, consistent with its remarkable triplet pairing state. This also applies to the alkali fullerides ($A_3C_{60}$'s), thus suggesting that they may also be exceptional, but the evidence (§2 here) shows that these materials follow the typical pattern of the other exotic superconductors. We must also emphasize that because of this clear difference between the uranium and ruthenium ions, and the active ions in most of the other exotic superconductors, the extra anomalies of the uranium heavy-fermion superconductors and $Sr_2RuO_4$ do not discredit the general picture presented in this paper. On the other hand, the fact that they share many characteristic features with other exotic superconductors (Brandow 1998) indicates that their pairing mechanism(s) must somehow be related to the main unconventional mechanism.

One of the criteria used to identify the presence or absence of gap nodes is the magnetic field dependence of the lowest-temperature specific heat [actually of $\gamma(H) = C(T,H)/T$ for $T \ll T_c$, ideally the $T \to 0$ limit], in the mixed state above $H_{c1}$. The standard theory of this field dependence (Volovik 1993, also the review of Yang and Lin 2001) has, however, been challenged by the observation of node-like behaviors in the $\gamma(H)$ of several materials which were thought to be conventional and thus presumed to have nearly isotropic gaps: $V_3Si$, $NbSe_2$, $CeRu_2$. This observation has led to an alternative explanation, namely, the vortex core shrinkage model, which appeals to a significant decrease of the vortex core radius with increasing magnetic field (the decrease being concentrated mainly at low fields) (Ramirez 1996, Sonier *et al*. 1999a). In reality, however, *these materials all have strong gap anisotropy*, as we demonstrate in this paper. It is now clear that these materials exhibit node-like behavior in their specific heat data because they do indeed have strong gap anisotropies, and, for $CeRu_2$, possibly even gap nodes in the best samples. Thus, what has previously been puzzling is actually the proper behavior, the behavior which should have been expected. We therefore ignore this model in the text, and relegate further discussion of this to Appendix B.

Exotic superconductors with cubic or nearly cubic crystal symmetry are discussed in §2, and planar (quasi-two-dimensional) exotics are examined in §3. A summary and concluding remarks are in §4. Appendix A discusses the Uemura trend and recent additions to the list of exotic superconductors. Appendix B shows that the vortex core shrinkage model is inadequate to explain the power-law field dependences of $T \ll T_c$ specific heat data, and that this model needs to be critically reexamined.

§2.   CUBIC MATERIALS

We begin this examination of exotic superconductors with cubic or nearly cubic examples, because these are often regarded as being conventional and are thus expected to have nearly isotropic s-wave gaps. (The recognized exception is the heavy-fermion $UBe_{13}$.) We now show that the gaps of some prominent examples are highly anisotropic, and that in some cases there may even be gap nodes.

2.1. *$Nb_3Sn$ and $V_3Si$*

The old and prominent A-15 materials $Nb_3Sn$ and $V_3Si$, which have relatively high $T_c$'s, have provided several kinds of evidence for strong gap anisotropy: (i) For $Nb_3Sn$, point-contact tunnelling into several inequivalent faces of a single complex crystal has produced different gap values, with $2\Delta/k_BT_c$ ranging from 1.0 to 2.8 (Hoffstein and Cohen 1969). (ii) A broad distribution of gap values, $2\Delta/k_BT_c$ = 1.0 - 3.8, has been observed for $V_3Si$ via far-infrared absorption (Tanner and Sievers 1973). (iii) The existence of strong gap anisotropy has been deduced for several A-15's ($Nb_3Ge$, $Nb_3Sn$, $V_3Si$, $V_3Ge$) by studying their $T_c$ reductions due to radiation damage (Farrell and Chandrasekhar 1977). This evidence deals only with Fermi-surface averages, and so does not provide maximum or minimum values for $2\Delta/k_BT_c$. (iv) A Raman-scattering study led to an estimate of 20% for the gap anisotropy of $Nb_3Sn$ (Dierker *et al.* 1983). This 20% was the difference between Fermi-surface averages in different symmetry channels ($E_g$ *vs.* $A_{1g}$), so the ratio $\Delta_{max}/\Delta_{min}$ can be expected to be much larger than 1.2.

## 2.2. *$K_3C_{60}$ and $Rb_3C_{60}$*

There is evidence suggesting considerable gap anisotropy in both $K_3C_{60}$ and $Rb_3C_{60}$. Infrared reflectivity data indicate approximately the BCS gap ratio (3.5) for both materials, $2\Delta/k_BT_c$ = 3.44-3.45 (Degiorgi *et al.* 1994, Degiorgi 1996), and likewise for the temperature dependences of NMR (nuclear magnetic resonance) (Stenger *et al.* 1995), μSR (muon spin relaxation) (Kiefl *et al.* 1993, MacFarlane *et al.* 1998), and the penetration depth (Neminsky *et al.* 1994), while point-contact tunnelling experiments have provided $2\Delta/k_BT_c \approx 5.3$ for both materials (Zhang *et al.* 1991a,b, Jess *et al.* 1994). A straightforward interpretation (Brandow 1998) is that the tunnelling value represents the gap *maximum* (see the tunnelling state density of Mahan 1989), while spin relaxation, penetration depth, and the onset of infrared absorption identify the gap *minimum*. This is consistent with other infrared experiments which reported mixtures or distributions of gap values, the observed gap ratio ranges being ~ 3-5 and ~ 2-5 (Rotter *et al.* 1992, FitzGerald *et al.* 1992) (although the authors doubted that these gap distributions were intrinsic). The infrared reflectivity data just mentioned also has features suggesting gap distributions.

Whether these distributions actually arise from gap anisotropies is, however, not at all clear. A reduced Hebel-Slichter peak has been observed (Kiefl *et al.* 1993, Sasaki *et al.* 1994, 1997, Stenger *et al.* 1995, MacFarlane *et al.* 1998), indicating or at least suggesting only minor anisotropy, and a combined study of tunnelling and optical transmission found a common gap magnitude from both of these techniques, $2\Delta/k_B T_c \approx 4.2$ (Koller *et al.* 1996). [Also, an early NMR study found strongly differing gap ratios $2\Delta/k_B T_c$: 3.0 for $K_3C_{60}$ and 4.1 for $Rb_3C_{60}$ (Tychko *et al.* 1992).] Various rationalizations can be offered to reconcile these data. There are sample questions of homogeneity of doping and stoichiometry at grain boundaries. The gap ratio of 4.2 in Koller *et al.* (1996) is a rough average of the values in the preceeding paragraph, suggesting angle-averaging of the gap by sample defects. The Hebel-Slichter peak also suggests this effect, although this peak could be due instead to the case of low-gap regions which are restricted to relatively small portions of the Fermi surface. Furthermore, the extraction of an effective $\Delta$ from low-temperature behavior has been argued to involve considerable subtlety and ambiguity (MacFarlane *et al.* 1998), at least potentially, so in principle any of the quoted $\Delta$'s from this approach might be criticized. Above all, however, we expect that the most serious problem may be the frozen-in rotational or orientational (merohedral) disorder of the $C_{60}$ molecules, which is known to produce strong variations in the various transfer integrals (Gelfand and Lu 1992a,b, Mazin *et al.* 1993, 1994). Such strong disorder may well go beyond making the gap essentially isotropic in k-space; it may also produce strong local variations of the gap in r-space, which may be sufficient to account for the observed gap distributions. It is therefore doubtful whether an "intrinsic" gap anisotropy can be determined for these materials. Nevertheless, all of these experiments agree that the gap is nodeless and therefore has s-wave-like symmetry.

2.3. *Cubic Laves-phase materials*

The cubic Laves-phase (C-15 structure) materials have not been directly confirmed as exotic. Nevertheless, HfV$_2$ (T$_c$ = 9.2K) has long been of interest because it shares several anomalous properties with A-15, Chevrel, and rhodium boride compounds: strong temperature dependence in normal-state magnetic susceptibility and Knight shift, large specific-heat γ, and a martensitic phase transition suggestive of strong electron-phonon coupling (Kishimoto *et al.* 1992, 2001). It also has a large and quite anomalous resistivity. These papers also present some gap-related data. The specific heat below T$_c$ shows clear T$^3$ dependence, suggesting that the gap vanishes at points on the Fermi surface. Below T$_c$ the 1/T$_1$ of NMR shows a very small Hebel-Slichter peak, and then a T$^5$ dependence which is clearly different from the activated behavior of an isotropic gap. This T$^5$ dependence likewise suggests that the gap vanishes at points, with a resulting state density ~ E$^2$. It could be, however, that the gap is instead merely *close to vanishing*, with $|\Delta_{min}| \ll \Delta_{max}$, where this $\Delta_{min}$ is either positive or negative. The Hebel-Slichter peak indicates that near the gap maximum the gap variation must be rather weak, so that the gap remains large over much of the Fermi surface.

The C-15 materials CeCo$_2$ (T$_c$ = 1.4K) and CeRu$_2$ (T$_c$ = 6.2K) exhibit several of the typical exotic features (Brandow 1998).[6] The specific heat for CeCo$_2$ below T$_c$ has shown T$^2$ behavior, down to the lowest-T data at 0.12T$_c$ (Aoki *et al.* 1997a,b). This indicates a strongly anisotropic gap and a possibility of gap nodes. In contrast, in an NMR study the 1/T$_1$ showed a weak Hebel-Slichter peak and a lower-T behavior characteristic of an anisotropic gap without nodes (Ishida *et al.* 1997). However, the act of crushing the sample (necessary for this NMR experiment) has surely induced dislocations and accompanying strains, and these defects may have significantly reduced the gap anisotropy. [That this can be a significant effect is demonstrated by the case of UPt$_3$, where, in the initial study of its superconductivity (Stewart *et al.* 1984), the grinding of the sample (for NMR study) was found to totally suppress the superconductivity. Annealing then restored the superconductivity.]

There is also evidence for gap anisotropy in CeRu$_2$, namely, the observation of a finite but *relatively small* Hebel-Slichter peak in the $1/T_1$ of NQR (nuclear quadrupole resonance) data (Matsuda *et al.* 1995, Mukuda *et al.* 1998), and especially in the finding that a variety of impurity dopings all *increase* the magnitude of this peak (Mukuda *et al.* 1998), evidently because the doping reduces the gap anisotropy. Modelling of this change of the peak magnitude suggested that $\Delta_{min}/\Delta_{max} \approx 0.74$, but this estimate is unreliable because the actual angular dependence of the gap is unknown. The $\Delta_{min}/\Delta_{max}$ could be far smaller (even zero), if the low-$\Delta$ regions are confined to small portions of the Fermi surface. Further support for strong anisotropy comes from a specific heat study (Hedo *et al.* 1998) that found C/T at T = 0.5K to depend on magnetic field roughly as $H^{1/2}$. A detailed μSR study of this material (Kadono *et al.* 2001) has found a rough consistency with the specific heat study, but this μSR study is subject to problems described in Appendix B.

Other aspects of the specific heat and NQR data for CeRu$_2$ are unfortunately unclear about this issue. The specific heat data (at zero magnetic field) is claimed to be inconsistent, showing both power-law (Sereni *et al.* 1989) and conventional (Huxley *et al.* 1993, Hedo *et al.* 1998) temperature behavior. But this claim is itself problematic, for several reasons: (1) When plotted as C/T *vs.* $T^2$, the claimed power-law data of Sereni *et al.* (1989) actually looks fairly conventional,[7] *i.e.* activated, at the lowest temperatures shown, but with a relatively small effective $\Delta$ which would represent the minimum of an anisotropic gap. Also, inspection of the corresponding C/T *vs.* T plot (Sereni *et al.* 1989) shows that $C \approx -AT + BT^2$ between 2K and 6K, rather than $C \approx BT^2$ which was implied, so the claim of $T^2$ behavior in this range is misleading. There is thus no clear qualitative disagreement with Huxley *et al.* (1993) and Hedo *et al.* (1998). (2) The claims of conventional behavior, for specific heat (Huxley *et al.* 1993, Hedo *et al.* 1998) and also for $1/T_1$ of NQR (Matsuda *et al.* 1995, Mukuda *et al.* 1998), are not based on T → 0 data but on data typically with $T > T_c/4$, so the obtained effective gap values do not represent the $\Delta_{min}$ of an anisotropic gap. (3) Some of the specific heat data (Sereni *et al.* 1989, Huxley *et al.* 1993, Hedo *et al.* 1998) show residual linear-in-T components at the lowest temperatures, suggesting normal-state inclusions, so a full agreement between these experiments should not be expected.

*2.4. BKBO*

In contrast to the preceeding examples, the existing evidence for the cubic perovskite BKBO ($Ba_{1-x}K_xBiO_3$) shows an essentially isotropic gap. Tunnelling data have shown very good agreement with BCS (isotropic gap) modelling (Sato *et al.* 1990, Huang *et al.* 1990, Sharifi *et al.* 1991, Zasadzinski *et al.* 1991, Kussmaul *et al.* 1993, Kosugi *et al.* 1994). The $T \ll T_c$ data for penetration depth (Pambianchi *et al.* 1994) and specific heat (Woodfield *et al.* 1999) show isotropic-gap behavior, and the optical conductivity of BKBO also looks quite conventional (Puchkov *et al.* 1994, Timusk 1999). BKBO thus appears "anomalously conventional" in this respect. This conclusion is also supported or at least suggested by the band-theoretic finding of a nearly spherical Fermi surface (Hamada *et al.* 1989).

It is not clear, however, whether this result is intrinsic or extrinsic. The samples with the highest $T_c$'s are at the edge of a compositional metal-insulator boundary. Because of fluctuations in the local dopant (potassium) density it is likely that the highest $T_c$ samples are electronically inhomogeneous, with metallic (superconducting) and insulating regions coexisting on a microscopic scale. Indeed there is considerable evidence for this (Hellman and Hartford 1993, Schmidbauer *et al.* 1994, Szabó *et al.* 1994, Misra *et al.* 1996, Yamato 1996, Zakharov *et al.* 1997). It is thus quite possible that local inhomogeneity will cause strong scattering within the majority metallic phase, and may thereby angle-average the gap sufficiently to provide a nearly isotropic behavior, even if the intrinsic gap were to have a significant anisotropy. This issue should be clarified by more careful study of samples farther from the macroscopic phase boundary.

2.5. *Other materials*

There are several other cubic or nearly cubic materials among the confirmed exotic superconductors (footnote 2 and Appendix A): Chevrel materials, $LiTi_2O_4$, silicon clathrates, the pyrochlore $Cd_2Re_2O_7$, and the highly anomalous $UBe_{13}$. For Chevrel materials there is tunnelling data showing a fully-gapped conductance form (Poppe and Wühl 1981), and also a clear activated behavior in NMR data, the latter remarkably extending over four decades (Kitaoka *et al.* 1992). For polycrystal $LiTi_2O_4$ there is some evidence for a rather isotropic gap (see Annett 1999), but since there is no single-crystal data it is possible that the intrinsic gap has significant anisotropy. For the recently discovered pyrochlore $Cd_2Re_2O_7$[8] there is NMR data (Vyaselev *et al.* 2002) showing a prominent Hebel-Slichter peak followed by exponential suppression of $1/T_1$ at lower T, which indicate an s-wave gap without much anisotropy. Muon spin relaxation studies of this material have also found approximately isotropic gap behavior, although with a hint of some anisotropy (Kadono *et al.* 2002, Lumsden *et al.* 2002). We are not aware of any gap-form evidence for the silicon clathrates.

The "obvious candidate" cubic materials include NbN and the recently discovered $MgCNi_3$ with anti-perovskite structure (the Ni's occupy the usual oxygen sites) (He *et al.* 2001, Li *et al.* 2001a). For NbN thin films, good isotropic-gap behavior is shown by tunnelling (Kashiwaya *et al.* 1991) and penetration depth data (Pambianchi *et al.* 1994, Komiyama *et al.* 1996, Lamura *et al.* 2002). For the anti-perovskite there is also NMR data (Singer *et al.* 2001) showing a prominent Hebel-Slichter peak followed by exponential suppression of $1/T_1$ at lower T, indicating an s-wave gap without much anisotropy. The temperature dependence of the specific heat shows isotropic-gap behavior, although there is some disagreement about the γ(H) behavior (Lin *et al.* 2003, Wälte *et al.* 2002). Surprisingly, however, point-contact tunnelling into $MgCNi_3$ has shown a very prominent zero-bias peak anomaly (Mao *et al.* 2001), a feature usually interpreted as a signature for gap nodes. There is a proposal to reconcile these contrasting features (nodeless gap *vs*. zero-bias anomaly) by means of a multicomponent gap, where each component is associated with a separate Fermi surface sheet and has nodeless s-wave character, but where there are phase differences between the different sheets (Voelker and Sigrist 2002). An alternative possibility is an effect of surface impurity or defect states (Samokhin and Walker 2001). We must point out that there is also a more straightforward possibility. Quite recent data (Prozorov *et al.* 2003) shows power-law ($T^2$) variation of the penetration depth for T less than or about $T_c/4$, which would imply either line nodes modified by defect scattering, or point nodes. In view of the preceeding nodeless behavior, this is evidence for strong sample dependence of the gap anisotropy. This therefore implies a strongly anisotropic *but s-like* intrinsic gap form. As we are now demonstrating, this is the typical gap character of an exotic superconductor. Other typical exotic fetures of this material are a "strange" chemical formula (with a nonmetal, carbon, and with Ni ions which should carry a substantial Hubbard U interaction), strong type-II character (κ ~ 46), and a small coherence length (ξ ≈ 46Å) (Mao *et al.* 2003).

Just as for $MgCNi_3$, it is possible that future improved samples of any of these other materials may provide evidence for anisotropy. This remark is not intended to minimize the possibility or likelihood that there are genuine (intrinsic) exceptions to the rule of strong gap anisotropy; we expect that such exceptions may well exist. The present point is simply that for a number of other materials (BKBO, $Cd_2Re_2O_7$, $LiTi_2O_4$, silicon clathrates, NbN) the absence of a significant *intrinsic* anisotropy has not been confirmed beyond reasonable doubt.

§ 3.  LAYERED MATERIALS

Among the layered or planar (quasi-two-dimensional) exotic materials there is much further evidence for a general anisotropic-s gap character. Much of this evidence comes from pairs of quite similar materials which *do* and *do not* exhibit gap nodes, and even from apparently equivalent samples which differ in this respect. The underlying argument was described in the Introduction, namely, that nodes can easily be removed from an anisotropic-s gap by reducing the gap anisotropy, in contrast to the cases of other gap symmetries. This argument helps to demonstrate or confirm that a variety of planar-material families all have s-like gap forms, with their gap nodes coming simply from stronger versions of the gap anisotropy which is already present in the cubic examples just discussed. This argument is also useful for near-node cases with $\Delta_{min} \ll \Delta_{max}$, where, due to the finiteness of the lowest available T, it is unclear whether true gap nodes are actually present. Although this type of evidence is not definitive, due to the conceptual possibility of a violation of time-reversal invariance, it is nevertheless strong evidence. This is generally consistent with the other evidence available.

It appears from the following examples (together with the cuprates), and §2, that gap nodes are more likely in quasi-two-dimensional materials, and there is a simple rationalization for this (§3.3 of Brandow 1998). The argument is based on the fact that in planar materials a weak c-axis coupling between planes leads to a nearly $k_z$-independent band structure. A strong coherence can therefore occur in the $k_z$ integration within the gap equation, which can thereby enhance (or fail to diminish) any underlying tendency for in-plane gap anisotropy.

3.1. *Borocarbides*

A broad variety of evidence demonstrates exotic character for the borocarbide superconductors, especially the prototype examples YNi$_2$B$_2$C and LuNi$_2$B$_2$C (Brandow 1998). We begin with the specific heat data. Several studies have found T$^3$ behavior at T << T$_c$, a feature which indicates gap nodes (*point* nodes) (Movshovich *et al.* 1994, Carter *et al.* 1994, Michor *et al.* 1995, Nohara *et al.* 1997, 2000), and this is also suggested by a study with a higher minimum temperature (Godart *et al.* 1995). There is also data showing crossover to activated behavior at lower temperatures (Hong *et al.* 1994, Izawa *et al.* 2001) (see footnote 7). This crossover implies absence of gap nodes, but a small $\Delta_{min}$ does indicate strong gap anisotropy. Another specific heat study found activated behavior over a broad temperature range (Schmiedeshoff *et al.* 2001), suggesting a nearly isotropic gap. This variation in the specific heat data strongly suggests change in anisotropy by variation in the sample quality, *i.e.* in the amount of defect scattering, which thereby indicates an anisotropic s-wave gap form.

The NMR 1/T$_1$ results are also variable. Most studies show absence of a Hebel-Slichter peak (Hanson *et al.* 1995, Kohara *et al.* 1995, Suh *et al.* 1996, Zheng *et al.* 1998, Iwamoto *et al.* 2000), although some studies have found such a peak (Saito *et al.* 1998, Mizuno *et al.* 1999, 2000). At lower T (down to ~4K) some of these studies found activated behavior (Hanson *et al.* 1995, Suh *et al.* 1996, Saito *et al.* 1998), while the others did not. Two studies down to quite low temperatures (<1K) found 1/T$_1$ ~ T (Zheng *et al.* 1998, Iwamoto *et al.* 2000), which they both attributed to gap nodes modified by impurities or disorder. One study found evidence for a significant fraction of the sample remaining normal (suggested to be due to vacancies and/or local interchanges of boron and carbon ions), with the normal signal component dominating 1/T$_1$ below 6K (Kohara *et al.* 1995). These variations of the 1/T$_1$ data likewise support a strongly anisotropic s-wave gap form for the intrinsic behavior, apparently with nodes.

Strong gap anisotropy is also demonstrated by Raman scattering (Yang *et al.* 2000), thermal conductivity (Boaknin *et al.* 2001), and magnetic field dependence of specific heat (Nohara *et al*. 1997, 1999, 2000, Izawa *et al*. 2001, Lipp *et al*. 2002), the latter two methods both suggesting gap nodes (*line* nodes). [However, the zero-field specific heat of Izawa *et al*. (2001) shows a very small region of activated behavior at the lowest T's (~2K), indicating a very small minimum gap.] A recent study of thermal conductivity has found evidence for *point* nodes which are oriented along the *x* and *y* axes of the Brillouin zone (Izawa *et al*. 2002a; see also Thalmeier and Maki 2002). This and the preceeding point-node evidence clearly favors an anisotropic-s gap form. On the other hand, a recent study of magnetic field orientation dependence of the specific heat has indicated *line* nodes, where the lines are parallel to the $k_z$ axis and pass through the $k_x$ and $k_y$ axes in the Brillouin zone (Park *et al*. 2002). We note that these point-node and line-node observations are compatible (compatible with different defect densities or scattering strengths) if the intrinsic gap form has elliptical gap nodes centered along the $k_x$ and $k_y$ axes, with the ellipses elongated in the $k_z$ direction.

    S-wave evidence from intentional reduction of gap anisotropy has been obtained for borocarbides in three different ways: (1) A suitable comparison material is $ThPt_2B_2C$, where one can reasonably expect the basic electronic parameters to be somewhat different. This material has a Hebel-Slichter peak in its NMR $1/T_1$, and at $T \ll T_c$ its $1/T_1$ is activated with a $2\Delta_{min}/k_B T_c \approx$ (BCS value of 3.5)/3 (Ikeda *et al*. 1996). The gap of this material is therefore clearly nodeless and less anisotropic. This comparison strongly suggests anisotropic gaps with s-like symmetry in all three of these materials ($ThPt_2B_2C$, $YNi_2B_2C$, and $LuNi_2B_2C$). (This argument is from §3.10 of Brandow 1998.) (2) Further s-wave evidence was obtained by a comparison of $YNi_2B_2C$ and $LuNi_2B_2C$ with $Y(Ni_{0.8}Pt_{0.2})_2B_2C$ (Nohara *et al*. 1999, 2000; see also Lipp *et al*. 2002). This work focussed on the magnetic field dependence of the specific heat γ at $T \ll T_c$, where the $Ni_2$ materials showed $H^{1/2}$ behavior (indicating or at least strongly suggesting line nodes), while the $(Ni_{0.8}Pt_{0.2})_2$ material showed linear-in-H behavior indicating a nodeless and nearly isotropic gap. [The γ(H) behavior of a sample with 5% Pt doping is also plotted in

Nohara *et al*. (2000). This has an apparent power-law exponent closer to 1.0 than to 0.5, showing that the anisotropy is already much reduced, and implying that the sample with 20% Pt doping should be nearly isotropic.] And plots of lnC(T) *vs*. $T_c/T$ for the Y materials (for zero and 20% Pt doping) have confirmed these results by showing change from $T^3$ to activated behavior for C(T) (Nohara *et al*. 2000). [In contrast, a comparison of YNi$_2$B$_2$C with Y(Ni$_{0.6}$Pt$_{0.4}$)$_2$B$_2$C via NMR found clear and comparable Hebel-Slichter peaks for both of these materials (Mizuno *et al*. 1999, 2000). We attribute this similarity to a high degree of disorder in both samples. The samples were arc-melted and then pulverized.] (3) In still another application of this argument, YNi$_2$B$_2$C and Y(Ni$_{0.8}$Pt$_{0.2}$)$_2$B$_2$C were compared in photoemission (non-angle-resolved) with high energy resolution (Yokoya *et al*. 2000). Modelling of the differences in the below-$T_c$ data of these materials indicated strong and weak s-wave gap anisotropies, respectively. These three separate comparisons, as well as the preceeding evidence, clearly demonstrate a strongly anisotropic s-wave gap form with nodes in the best samples.

As a final example of variability in the borocarbide gap anisotropy, we mention recent tunnelling data where the conductance plot is flat-bottomed (nodeless) and shows $\Delta_{min} \approx 0.4\Delta_{max}$, which is clearly an anisotropic s-wave result (Martínez-Samper *et al*. 2003).

Compared to the cuprates and the planar organics, the Ni borocarbides have only a weak *ab vs. c* anisotropy in $H_{c2}$ (Metlushko *et al*. 1997), so an assumption of cylindrical gap geometry [as in Eq. (1)] is unrealistic. Recent papers have modeled borocarbide data with a noncylindrical anisotropic-s gap form having point nodes (Izawa *et al*. 2002a, Maki *et al*. 2002, Thalmeier and Maki 2003, Yuan and Thalmeier 2003), as well as with the cylindrical form in Eq. (1) (Lee and Choi 2002). The Fermi surfaces of these materials are actually quite complicated -- in each case the Fermi surface involves several different (inequivalent) sheets -- so a single-sheet representation may be an oversimplification.

3.2. *Planar organics*

The layered or planar organic superconductors also present a striking case for the anisotropy reduction argument. (We focus mainly on the κ-(ET)$_2$X family, which has the highest-$T_c$ examples.) The gap-node aspect of these materials has long been and still continues to seem paradoxical. Many experiments have shown behavior indicating a nodeless and perhaps fairly isotropic gap, while many other experiments have demonstrated gap nodes. For example, NMR data for κ-(ET)$_2$Cu[N(CN)$_2$]Br from several laboratories (De Soto *et al.* 1995, Mayaffre *et al.* 1995, Kanoda *et al.* 1996) show clear evidence for gap nodes, via $1/T_1 \propto T^3$ at $T \ll T_c$ (these NMR samples were *not* crushed), and also the absence of a Hebel-Slichter peak, whereas recent specific heat data for this same material (and others) just as clearly demonstrates the *absence* of gap nodes (Elsinger *et al.* 2000 and references therein, Müller *et al.* 2002a). An earlier specific-heat study of this material did, however, find evidence of gap (line) nodes (Nakazawa and Kanoda 1997). This dichotomy is also found in the many studies of temperature dependence of the London penetration depth, which is well known for providing both power-law (node) behavior (Kanoda *et al.* 1990, 1993, Le *et al.* 1992, Achkir *et al.* 1993, Tsubokura *et al.* 1995, Carrington *et al.* 1999, Pinteric *et al.* 2000, Pratt *et al.* 2001) and activated (nodeless) behavior (Harshman *et al.* 1990, 1994, Klein *et al.* 1991, Lang *et al.* 1992a,b, 1994, Dressel *et al.* 1993, 1994). Several other types of experiments have indicated gap nodes: thermal conductivity (Belin *et al.* 1998, 1999, Behnia *et al.* 1999, Izawa *et al.* 2002b), millimeter-wave cavity absorption (Schrama *et al.* 1999, Schrama and Singleton 2001), and tunnelling (Arai *et al.* 2001). Earlier tunnelling work by the same team (of Arai *et al.* 2001) had, however, found gap anisotropy without nodes (Nomura *et al.* 1995, Ichimura *et al.* 1997).

Altogether, there are so many experiments showing or suggesting gap nodes, and so many others showing absence of nodes, it now seems inescapable that at least some of the results of both types must be correct. It is no longer tenable to presume (as some researchers have) that because some results are correct the contrary results must somehow be wrong. The most reasonable conclusion is that the degree of gap anisotropy in these materials is very sensitive to unintentional perturbations, by means of which nodes can easily be removed. There is indeed a very plausible source for such perturbations, in the various sample cooling histories. There may well be insufficient recognition that these materials must be slowly cooled, or annealed, over a temperature range ~100K-60K, in order to avoid or at least significantly reduce disorder among the ethylene groups in the ET molecules (*e.g.* Pouget 1993, Aburto *et al.* 1998, Su *et al.* 1998, Tanatar *et al.* 1999a,b, 2000a,b, Akutsu *et al.* 2002, Müller *et al.* 2002b, Pinteric *et al.* 2002).[9] (This disorder distorts the stacking of the ET molecules, and thereby perturbs the transfer integrals between these molecules. This is rather like the disorder in fullerene superconductors mentioned in §2, but this evidently has weaker effect here.) Other sample defects may also be significant here [Pinteric *et al.* 2002; see also comments on κ-$(ET)_2Cu(NCS)_2$ samples in Dressel *et al.* 1994]. Thus, the extensive bimodal character of the data is strong evidence for an anisotropic-s type of gap.

Another application of this anisotropy-reduction argument is to compare the above NMR data with the corresponding data for κ-$(MDT-TTF)_2AuI_2$. In the latter material the unsymmetrical form of the MDT-TTF molecule (Papavassiliou *et al.* 1988) can reasonably be expected to provide strong disorder within the crystal. A prominent Hebel-Slichter peak was found (Kobayashi *et al.* 1995), in contrast to the previously quoted data, indicating an s-wave gap without strong anisotropy.[10] Also, a later specific heat study found an essentially isotropic gap behavior (Tsubokura *et al.* 1997). [On the other hand, a tunnelling study of this material found evidence for strong gap anisotropy (Ichimura *et al.* 1999), which we presume was due to better ordering through careful sample preparation.] These results are further support for a general s-wave symmetry throughout the planar organics. And, surprisingly, except for a less-detailed version of this argument in Brandow (1998), we are not aware of any previous suggestion of an anisotropic s-wave gap form as a resolution for the data dichotomy of these materials.

With a band structure nearly independent of $k_z$, and very weak pair coupling between the planes (presumably by Josephson tunnelling), the gaps of these materials can reasonably be expected to have a form similar to Eq. (1), although with some orthorhombic distortion. A rough description of the Fermi surface is that this consists of a circle (or cylinder), for which Eq. (1) could be appropriate. This circle extends outside of the Brillouin zone, however, so the extended-zone description consists of overlapping circles, and the intersection regions show avoided crossings (*e.g.* Caulfield *et al.* 1995, Ching *et al.* 1997). The consequences of these avoided-crossing regions for a simple description of the gap form are unclear. Since the location of the gap nodes is not our present concern, we merely note that the relevant (and conflicting) evidence has recently been reviewed (Singleton and Mielke 2002).

3.3. *NbSe$_2$*

Niobium diselenide is still often regarded as conventional (purely phonon driven), and is thus assumed to have a fairly isotropic gap, even though this material has been shown to be exotic.[2] Although there is data suggestive of gap nodes (Nohara *et al.* 1999, 2000), there is more quantitative evidence from tunnelling, specific heat, and NQR data that NbSe$_2$ does not actually have nodes but instead has an anisotropic-s gap with $\Delta_{min} \approx \Delta_{max}/2$ (Hess *et al.* 1991, Sanchez *et al.* 1995, Ishida *et al.* 1996, Hayashi *et al.* 1997). This may be the main reason why the apparent specific-heat "signature" for gap nodes is noticeably weak in this material, *i.e.*, why the specific heat H dependence observed for pure NbSe$_2$ is intermediate between the $H^{0.5}$ (gap with line nodes) and $H^{1.0}$ (isotropic gap) forms (Nohara *et al.* 1999, 2000). [The finite temperature at which this data was taken apparently also helps to obscure the difference between true nodes and strong anisotropy with a small finite $\Delta_{min}$, in effect interpolating between their ideal (T = 0) behaviors.]

In contrast, there is other specific heat data (Sonier *et al*. 1999a) and μSR data (Sonier *et al*. 1997) which have been fitted to an *isotropic* s-wave model with a vortex core radius (~ coherence length) which *rapidly diminishes* under an applied magnetic field (the vortex core shrinkage model). Since these fits have ignored the known strong gap anisotropy, the significance of the model for this material is unclear. We comment further on this model in Appendix B.

3.4. *MgB$_2$*

The question of a possible unconventional contribution to the pair interaction in MgB$_2$ is still being debated. Just as for many of the confirmed exotic superconductors, there have been band-theoretic calculations of the electron-phonon coupling λ which found values large enough (~0.7-0.9) to possibly account for the surprisingly large T$_c$ of 39K. On the other hand, several features have been emphasized as problematic for totally phononic pairing: small boron, magnesium, and total isotope effects (Hinks *et al*. 2001, Knigavko and Marsiglio 2001, 2002, Cappelluti *et al*. 2002, Zhao 2002), and evidence for a very small *transport* electron-phonon coupling (λ$_{tr}$ ≈ 0.15) and an associated large discrepancy between the measured (1.7eV) and conventional band-theoretic (≈7eV) values of the Drude plasma frequency (Tu *et al*. 2001, Marsiglio 2001). All of these problems are now claimed to be resolved by taking proper account of the unusual band structure of this material (the quite different characters and phonon couplings for the σ and π bands, and an extremely weak defect scattering between these bands), together with a strong anisotropy of the electron-phonon coupling (Mazin *et al*. 2002, Choi *et al*. 2002a,b, Maksimov *et al*. 2002, Marsiglio 2002).

In spite of this apparent success, the claim that the high T$_c$ is due to the phonons alone must be considered suspect. This is because the standard band-theoretic methods of calculating the electron-phonon λ do not take account of the "strong correlations" induced by the short-ranged screened Coulomb interaction – the Hubbard interaction U.

Several studies have shown that the U-induced correlations should reduce the effective coupling strength $\lambda$.[11] It is therefore appropriate to ask whether $MgB_2$ has any of the anomalous features typical for the exotic superconductors, features which would suggest an unconventional contribution in the pair interaction. We note that $MgB_2$ has such typical exotic features as strong type-II character ($\kappa_c = \lambda_{ab}/\xi_{ab} \approx 15$) and a relatively short coherence length ($\xi_{ab} \approx 65\text{-}70\text{Å}$) (de Lima *et al*. 2001a,b, Xu *et al*. 2001). (The $\lambda_{ab} \approx 1000\text{Å}$ used here is taken from Niedermayer *et al*. 2002.) There is also low-power-law resistivity.[12] We also find it intriguing that the crystal structure has the Mg ions rather isolated from each other by cages of boron ions, similar in this regard to the structures of the exotic materials $NbSe_2$, $U_6Fe$, the silicon clathrates, and BKBO.[13] (The $CuO_2$ planes of the superconducting cuprates are also examples of this geometric feature.)

A μSR study indicates that $MgB_2$ is probably exotic. This is discussed in Appendix A, where it is shown that $MgB_2$ is "less exotic" than most of the previously confirmed exotic materials, in the sense of lying below the main trend in the Uemura plot. But it is "more exotic" in this sense than $YNi_2B_2C$, since the latter lies still farther from this main trend (see Appendix A). Nevertheless, $YNi_2B_2C$ clearly exhibits a number of the characteristic exotic features (Brandow 1998), including the strong gap anisotropy discussed above. We suggest that being near or somewhat beyond the previous lower edge of the Uemura trend is due, in these cases, to the conventional phonon pairing contribution being relatively quite important, that is, important in the determination of $T_c$. From these considerations we conclude that $MgB_2$ is very likely exotic, in the sense of having an unconventional contribution in its pair interaction. This should also be significant for properties other than the remarkably high $T_c$, for example in providing the short coherence length.

The discussion of gap-related features has mainly focussed on the $H_{c2}$ anisotropy between the *ab* planes and the *c* axis, and on the evidence for two gaps (different gaps for the σ and π bands). Although there could also be considerable gap anisotropy *within* the *ab* planes, we are not aware of any direct evidence for this. Most of the T << $T_c$ data displays activated behavior demonstrating absence of gap nodes. Consistent with this, simple prolate and oblate spheroidal gap models have been proposed (Haas and Maki 2001, Seneor *et al.* 2001, Posazhennikova *et al.* 2002, Mishonov *et al.* 2002, Dahm *et al.* 2002a). On the other hand, we note that there are at least six low-T experiments showing or at least suggesting power-law behaviors which seem to indicate gap nodes (Panagopoulos *et al.* 2001, Li *et al.* 2001b, Wang *et al.* 2001, Zhukov *et al.* 2001, Pronin *et al.* 2001, Ohishi *et al.* 2003). [Admittedly, the lowest temperatures in some of these experiments are too high to be convincing. There is also strongly nonlinear γ(H) data (Wang *et al.* 2001, Yang *et al.* 2001b) which might be thought to indicate gap nodes, but in this case the nonlinearity has been shown to arise from the multiple Fermi surface sheets (multiple conduction bands) with quite different gap magnitudes (Nakai *et al.* 2002). On the other hand, although the power-law T dependence of λ(T) from Panagopoulos *et al.* (2001) has been argued to be spurious by Niedermayer *et al.* (2002), the latter authors did not consider the possibility of a significant sample difference.] In common with the preceeding materials, this power-law evidence suggests an s-wave gap form with intrinsic nodes. If this is correct, this implies that most of the samples to date have had too many defects to preserve these nodes. This apparent gap-node evidence clearly deserves further study.

We are not aware of any gap-form evidence for the remaining exotic materials with layered crystal structures – intercalated $Hf_2N_2Cl_2$ and $Y_2C_2(Br,I)_2$ (see Appendix A) – apart from heavy-fermion and uranium materials which we are not considering.

§ 4.  CONCLUDING REMARKS

We have shown that the existing gap-form evidence demonstrates strong gap anisotropy in several cubic or nearly-cubic examples of exotic superconductors, where the conventional expectation is for nearly isotropic gaps. And for planar examples the evidence shows anisotropy often strong enough to provide gap nodes, while the gaps nevertheless still have overall s-wave-like symmetry consistent with the cubic examples. (Notably, this gap behavior is found for the controversial cases of the planar organics and the borocarbides $YNi_2B_2C$ and $LuNi_2B_2C$.) Stated somewhat differently, we have shown that: (1) For *most* of the exotic superconductors[14] the gap symmetry is s-wave-like, meaning that the gap exhibits the full point-group symmetry of the crystal. (This may well also include the cuprate superconductors; see below.) (2) Strong gap anisotropy is very common in the exotics, although probably not universal. The anisotropy is typically stronger in the planar materials, where gap nodes are often possible for samples of sufficiently high quality, and a plausible reason for this was given. (BKBO and the "other materials" at the end of Sec. II (except for $MgCNi_3$) are good candidates for weak anisotropy, but it is not clear yet whether any of their apparently isotropic behaviors are intrinsic.)

Although skeptics might argue that for particular materials the evidence is not definitive, we emphasize that the present conclusions are the most straightforward and therefore the most reasonable interpretations of the data. The overall consistency of these results is also significant. Together with the many other phenomenological similarities throughout the exotic superconductors (Brandow 1998, 2000), this strongly supports the concept of a shared unconventional pairing mechanism.

These conclusions represent a middle course between two extreme but common views that are now shown to be false, namely, "gap nodes mean a non-s gap symmetry", and "absence of gap nodes means conventional purely-phonon-driven superconductivity". The latter view is disproved by the Uemura relation and the many other typical exotic features of these materials, as well as by the absence of comparably strong gap anisotropy among the purely phonon driven superconductors. The s-wave character is of course symmetry-consistent with the conventional phonon mechanism, and it allows the latter to contribute substantially (indeed there is usually evidence for this), but the typically strong gap anisotropy is one of the many indications that there is also a significant unconventional contribution to the pair interaction. Conversely, this typical cooperation between the conventional and unconventional mechanisms is further evidence for the general s-wave gap character in the exotics. We also note that throughout the exotics this synergy is consistent with the considerable deviations from a strict $T_c \propto \lambda_L^{-2}$ relation, and with the considerable range of $\xi/a$ values ($\xi$ = coherence length, $a$ = relevant lattice parameter). It thus appears that the relative strength of the new mechanism, as compared to the conventional one, varies widely from one material to another (and varies even between different samples of the "same" material with various dopings or stoichiometries).

Elsewhere, we have presented and reviewed much evidence for a nodal s-wave gap form in the hole-doped cuprates, and have also shown that the evidence for the d-wave gap form is considerably weaker than is generally believed (Brandow 2002).[15] Strong evidence for s-wave gaps in *electron-doped* cuprates is of course well known, but there is also recent data indicating strong gap anisotropy and sometimes even gap nodes in these materials (Brandow 2002). There is also more data supporting this picture of a general anisotropic-s gap form. Some of this data shows both node behavior (Ku *et al*. 2001) and nodeless behavior (Skinta *et al*. 2002a) for optimally doped films of the electron-doped material $Pr_{1-x}Ce_xCuO_4$, and node evidence for an optimally doped (x = 0.15) single crystal (Balci *et al*. 2002). Other data shows nodes for underdoped films together with the absence of nodes for overdoped films of this material, and likewise for the lanthanum analog of this material (Biswas *et al*. 2002, Skinta *et al*. 2002b; see also Zheng *et al*. 2002). But instead of characterizing the latter electron-doped cuprate results as "a *d* to *s* transition", as in these papers, it is far more straightforward and therefore more reasonable to interpret this as simply a change in the degree of anisotropy within the s-wave gap form of Eq. (1). [There is also recent data claiming a doping-dependent change of symmetry for the hole-doped cuprate YBCO ($YBa_2Cu_3O_{7-\delta}$) (Yeh *et al*. 2001, 2003; see also Yeh *et al*. 2002), and a number of other reports show evidence for node-nodeless transitions in hole-doped cuprates, as functions of T and/or H and/or doping (Krishana *et al*. 1997, Ong *et al*. 1997, 1999, Movshovich *et al*. 1998, Aubin *et al*. 1999, Sonier *et al*. 1999b, Vobornik *et al*. 1999, Ando *et al*. 2000, 2002, Gonnelli *et al*. 2001, Dagan and Deutscher 2001). These observations may also be consistent with the present picture.]

For another electron-doped material, $Nd_{1-x}Ce_xCuO_4$, there is also angle-resolved photoemission data (Sato *et al*. 2001) and optical data (Yanagisawa *et al*. 2001) which indicate gap nodes. There is also recent tunnelling data (both c-axis and randomly oriented) for the electron-doped infinite-layer material $Sr_{0.9}La_{0.1}CuO_2$, which is the other cuprate crystal type with electron-doped superconductivity (Chen *et al*. 2002, Yeh *et al*. 2002b). In both cases (c-axis and randomly oriented) the observed conductance form is close to that of a d-wave gap, but is more rounded (U-like) at zero voltage. These features suggest the gap form (1) with strong anisotropy, but without nodes. This is also consistent with NMR studies of this material, which did not find a Hebel-Slichter peak (Imai *et al*. 1995), and which found anisotropy weaker than in a conventional d-wave form (Williams *et al*. 2002).

Altogether, it appears that defects are less effective in removing gap nodes from the hole-doped cuprates than from the electron-doped cuprates and other exotic superconductors. This could be due simply to the typically larger energy scale of the pairing gaps in the hole-dopped cuprates.

In common with the other exotics, in the hole-doped cuprates there is also evidence for a synergy between a new pairing mechanism and the conventional phonon mechanism, and indeed this synergy is part of the s-wave evidence presented for the cuprates. [Point (4) of §2 of Brandow (2002) focussed on the smooth and strong monotonic increase in the oxygen isotope shift α for YBCO ($YBa_2Cu_3O_{7-\delta}$) when its $T_c$ is reduced by doping. It is also suggestive that among the various cuprate superconductors the typically large gap ratios $2\Delta/k_BT_c$ are increasing with increasing $T_c$; these ratio values follow a linear variation with $T_c$ which extrapolates back to the BCS value (3.5) at $T_c = 0$ (Wei *et al*. 1998).] It is therefore quite possible and even likely that the present gap-form systematics holds also for the cuprates (both hole-doped and electron-doped), thus providing a general consistency of the present features throughout most[14] of the exotic superconductors.

At the risk of belaboring this discussion, we mention a further implication of the present results. The conclusion that most of the non-cuprate exotics have s-like gap forms can be viewed as a further indication that the cuprates should have such a gap form. (This is an addition to the s-wave arguments of Brandow 2002.) This argument is significant because of the many other similarities throughout the exotic superconductors, including the Uemura relation, which strongly suggest a shared or common new pairing mechanism. This is the converse of the search for d-wave gaps in other exotics, which has been motivated mainly by the apparent d-wave conclusion for the cuprates.

We close by mentioning some further consequences of the present results: (1) It is now clear that the vortex core shrinkage model, which is partly based on the assumed gap isotropy of $V_3Si$, $NbSe_2$, and $CeRu_2$, is inadequately justified and needs to be carefully reexamined. (See Appendix B.) This illustrates that failure to recognize strong gap anisotropy can have serious consequences. (2) The conclusions of general s-wave character and typically strong gap anisotropy are significant constraints for the new or unconventional pairing mechanism, the mechanism which is apparently active throughout most of the exotic superconductors and responsible for their exoticity. We have shown that a valence-fluctuation mechanism does provide these features, and that this is also consistent with many other characteristic features of the exotic superconductors (Brandow 1994, 2000).

ACKNOWLEDGEMENTS

I thank D. F. Agterberg, L. N. Bulaevskii, M. J. Graf, J.-Y. Lin, F. Marsiglio, A. P. Ramirez, J. E. Sonier, I. Vekhter, and G.-m. Zhao for helpful discussions. This work was partially supported by the U. S. Department of Energy.

APPENDIX A: Recently Established Exotic Superconductors, and the Uemura Plot

In addition to the older and more widely recognized exotic superconductors listed in footnote 2 (detailed references in Brandow 1998), several more examples have recently been established. We note at the outset that to decide whether a given material is exotic or not, we must be guided primarily by the original criterion of Uemura and co-workers, that the data point for the μSR relaxation rate $\sigma \propto \lambda_L^{-2}$ lies within or near to the established Uemura trend, the band of data points where $T_c$ is approximately proportional to $\lambda_L^{-2}$. The basis for this trend is still poorly understood (several interpretations are discussed in Brandow 1998, 2000), and this problem needs to be approached both theoretically and empirically. Our present orientation is purely empirical.

The materials recently shown to lie within this trend, or at least near to this, are: (a) intercalated $Hf_2N_2Cl_2$ (Uemura *et al.* 2000), (b) the silicon clathrate $Ba_8Si_{46}$ (Gat *et al.* 2000), (c) $Y_2C_2(Br,I)_2$ (Henn *et al.* 2000), (d) the pyrochlore $Cd_2Re_2O_7$ (footnote 8), and (e) $MgB_2$ (Niedermayer *et al.* 2002). The papers on $Cd_2Re_2O_7$ and $MgB_2$ did not place their results on the Uemura plot. We now do this in the following manner: In the older Uemura plots (of $T_c$ *vs*. $\sigma \propto \lambda_L^{-2}$; see footnote 1), the band of the established exotic data points is seen to extend from the "Uemura line" (the upper margin of the Uemura band) to a value of σ (the μSR relaxation rate) about five times larger than this (for a given $T_c$ value). This factor of five is obtained from the data points for $V_3Si$, $NbSe_2$, and $U_6Fe$, which lie along the apparent bottom of the Uemura band. The σ for $Cd_2Re_2O_7$ is 0.085-0.10 $\mu s^{-1}$, which is to be compared with $\sigma \approx 0.025$ $\mu s^{-1}$ for the Uemura line at $T_c = 1K$. The ratio of less than or about 4 here leaves this data point clearly within the Uemura band, so by the usual criterion this material definitely qualifies as exotic. The σ for $MgB_2$ is about 7.9 $\mu s^{-1}$, which is to be compared with $\sigma \approx 1.0$ $\mu s^{-1}$ for the Uemura line at $T_c = 39$K. This point therefore lies beyond the previous apparent band margin, and $MgB_2$ can thus be said to be only marginally exotic.

A further comparison is appropriate at this point. We note that for $YNi_2B_2C$ the μSR-determined $\lambda_L = 1030$Å (Cywinski *et al*. 1994) is essentially the same as for $MgB_2$ (Niedermayer *et al*. 2002), so the corresponding σ must also be nearly the same, whereas for the $T_c$ of 15K the Uemura line value is 0.38 $\mu s^{-1}$. The σ for $YNi_2B_2C$ thus exceeds this Uemura-line value by a factor of about 21, which puts this material rather far beyond the previous apparent Uemura band margin. Nevertheless, $YNi_2B_2C$ clearly shares a number of the characteristic features of the other exotics (see Brandow 1998 and the presently discussed strong gap anisotropy with nodes). Because of the inadequate understanding of the Uemura plot, and especially of the boundary between the exotic and conventional superconductors, we must be guided here by the phenomenology of these characteristic features. Therefore, since $YNi_2B_2C$ is exotic in some degree (according to the latter evidence), we are led to conclude that $MgB_2$ is probably also exotic.

We interpret this low or marginal exoticity, for materials near or below the apparent bottom of the Uemura band, as probably meaning that the conventional pairing mechanism is playing a major role in determining the actual $T_c$, as compared to the unconventional contribution. (But this does not necessarily mean that the influence of the unconventional mechanism must be small for all other properties; for example the coherence lengths in $MgB_2$ and the borocarbides are quite short compared to conventional superconductors.) This marginality contrasts with $Hf_2N_2Cl_2$ (intercalated) and $Y_2C_2(Br,I)_2$, for which the data points are on or close to the Uemura line. The clathrate $Ba_8Si_{46}$ plots near but to the left of $NbSe_2$, thus showing relatively low exoticity.

APPENDIX B: Comments on the vortex core shrinkage model

As noted in the Introduction, an alternative explanation has been proposed for the approximate $H^{1/2}$ dependence, or downwards curvature in the H dependence, of the specific heats of various materials at $T \ll T_c$ (Ramirez 1996, Sonier *et al.* 1999a, Kadono *et al.* 2001). This is based on a shrinkage of the vortex cores in the mixed state, shrinkage which involves a rapid initial decrease of the core radius under applied magnetic field (Sonier *et al.* 1997, Kadono *et al.* 2001). (This shrinkage is attributed to vortex-vortex interactions. For the connection between core radius and specific heat, see Yang and Lin 2001.) It is claimed that this shrinkage has been confirmed rather directly, by fitting the Fourier transform of the local magnetic field distribution B(r) obtained from μSR data, using a theoretical form for the flux-lattice state (Sonier *et al.* 1997).

A major motivation for applying this model has been the finding that γ(H) varies roughly as $H^{1/2}$ for several supposedly conventional s-wave materials ($V_3Si$, $CeRu_2$, $NbSe_2$) whose pairing gaps were presumed to be nearly isotropic. (But see below for $V_3Si$.) The borocarbides $YNi_2B_2C$ and $LuNi_2B_2C$ have sometimes also been included in this context (*e.g.* Ohishi *et al.* 2002). The ample evidence here in §§ 2, 3 shows that *these materials actually all have strong gap anisotropy*, to the extent that the best samples of the borocarbides have gap nodes, and perhaps likewise for $CeRu_2$. The conspicuous departures from γ(H) ∝ $H^{1.0}$ (expected for an isotropic gap) are therefore quite reasonable and do not need an alternative explanation. (See below for $V_3Si$.) Core shrinkage has been considered as a possible explanation for the strong doping sensitivity of the γ(H) power law (Nohara *et al.* 1999), but no clear mechanism for this was found.

It is significant here that the main experimental evidence for vortex core shrinkage in these materials (shrinkage under applied fields) is not conclusive. This is because the model used to fit the μSR data is based on the assumption of an isotropic gap, and it is therefore quite possible that at least some of the apparent shrinkage may be an artifact arising from gap anisotropy. [This applies also to μSR studies (Sonier *et al.* 1999b,c) of a cuprate superconductor which is well known to have gap nodes. This seems to support the conjecture.] There is some data in poor agreement with this model (Miller *et al.* 2000,

Kadono *et al*. 2001), and also some evidence directly challenging this model (Izawa *et al*. 2001). For all of these reasons this model needs to be critically reexamined.

Some confusion has resulted from the paper of Ramirez (1994), on $\gamma(H)$ for $V_3Si$. This paper implied that the behavior of $\gamma(H)$ at small H (H ~ $H_{c1}$) would be relevant for the gap-form issue, if this small-H behavior were not confounded by vortex core shrinkage. This is consistent with the standard theory (Volovik 1993, Yang and Lin 2001), where only the larger-H (H >> $H_{c1}$) region of $\gamma(H)$ is relevant as gap-form evidence. This paper presented a much-quoted approximate $H^{1/2}$ [more correctly $(H - H_{c1})^{1/2}$] behavior, *but only for small H* ($H_{c1}$ < H less than or about $10H_{c1}$);[16] for larger H the $\gamma(H)$ form was found to be linear. (We conclude from this linear behavior that the gap in this $V_3Si$ sample had been made essentially isotropic by a substantial defect scattering.) Nevertheless, this paper has been widely quoted as showing a Volovik (1993) power law behavior of $H^{1/2}$ for $V_3Si$, and therefore as being a counterexample to the claim of $H^{1.0}$ behavior for nearly isotropic superconductors. This paper also mentioned downwards curvature in the $\gamma(H)$ of niobium (Ferreira da Silva *et al*. 1969) in this context, but this curvature is also irrelevant for the gap form. The Ginzburg-Landau $\kappa$ of niobium is so small (~1) that there is no $H_{c1}$ << H < $H_{c2}$ parameter region for this material (see Ferreira da Silva *et al*. 1969), so that a Volovik type of power law cannot be determined.

We are not arguing against vortex core shrinkage *per se*. There are theoretical calculations predicting this (Golubov and Hartmann 1994, Ichioka *et al*. 1999a,b), and to some extent the above-mentioned μSR evidence may be valid. However, the theoretically obtained power law exponents n, for $\gamma(H) \propto H^n$ (Ichioka *et al*. 1999a), are 0.67 (instead of 1.0) for an isotropic gap, and 0.41 (instead of 0.5) for a d-wave gap (and thus quite possibly for any other gap form with line nodes). This presents a problem because the majority of the available $\gamma(H)$ data favors the simple power law of 1.0 for the essentially isotropic case. (The difference between 0.41 and 0.5 is harder to distinguish experimentally.) There are some exceptional cases (*e.g.* Yang *et al*. 2001b, Wang *et al*.

2001, Schmiedeshoff et al. 2001, Lipp et al. 2002), but for these the smaller power-law exponents might possibly be due to having multiple conduction bands with different gap magnitudes (Nakai et al. 2002, Boaknin et al. 2002). (See also Amin et al. 2000, Dahm et al. 2002b, Kusunose et al. 2002.) The actual origin of these exceptions and the issue of possible discrepancies for single-band materials remain to be clarified.

FOOTNOTES

[1] See for example Uemura et al. (1991), Uemura and Luke (1993). Many other related papers are listed in Brandow (1998) and in Uemura (2003).

[2] The list of established exotic superconductors includes, in addition to the cuprates, examples from the families of bismuthates ($Ba_{1-x}K_xBiO_3$ or BKBO), alkali fullerides ($A_3C_{60}$'s), A-15 compounds, Chevrel compounds, organic materials (one-dimensional and two-dimensional), heavy-fermions and some related "almost heavy" materials ($U_6Fe$, $UPd_2Al_3$, $URu_2Si_2$, $U_2PtC_2$), and also $NbSe_2$ and $LiTi_2O_4$ (the first recognized high-$T_c$ oxide). References can be found in Brandow (1998), where it is also argued that several other materials should be considered exotic: the nickel borocarbides, the short-chain Chevrel analogs, the cubic Laves-phase compounds $CeRu_2$ and $CeCo_2$, and $Sr_2RuO_4$. See also recent additions to this list in the present Appendix A.

[3] For a brief summary of this report and its main implications, see §2 of Brandow (2000).

[4] See the compilations listed in Ref. 1 of Brandow (1998), and other compilations mentioned in Cava (1997).

[5] The electronic-parameter argument was used in §7.3 of Brandow (1994), and the disorder argument was used in §§ 3.6, 3.10 of Brandow (1998). The latter argument is implicit in Anderson (1959), and has been proposed a number of times, e.g. by Hotta

(1993), Norman (1994), Borkowski and Hirschfeld (1994), Fehrenbacher and Norman (1994), Kim and Nicol (1995), Pokrovsky and Pokrovsky (1995), Preosti and Musikar (1996), Yokoya *et al*. (2000).

[6] A further and striking feature is that the "magnetic resistivity" of $CeRu_2$, namely $\rho(CeRu_2) - \rho(LaRu_2)$, has a maximum near 100K and a logarithmic-in-T decrease above this (Nakama *et al.* 1995). This resistivity feature is found in a number of exotic superconductors, especially the planar organics (Brandow 1998, 2000).

[7] Compare for example with the specific heat of niobium, which is known to have only a small (few percent) gap anisotropy (Ferreira da Silva *et al*. 1969).

[8] Hanawa *et al*. (2001), Jin *et al*. (2001), Sakai *et al.* (2001), Hiroi and Hanawa (2002), Wang *et al.* (2002), Hiroi *et al*. (2002).

[9] We reject the conclusion in the above papers of Tanatar *et al*. that the resistance maximum (typically observed in these materials near 100K) is due entirely or even primarily to the ordering-disordering of the ethylene groups. An electronic (valence-fluctuation) origin for this maximum is proposed in §5.1 of Brandow (2000). Also in this mechanism, changes in the transfer integrals due to the ethylene-group ordering (via annealing) can be expected to strongly modify the resistance maximum.

[10] See for example the theoretical studies of Gabovich and Voitenko (1996), Ichinomiya and Yamada (1997).

[11] In the valence-fluctuation (Anderson lattice model) context this has been shown by Kim *et al*. (1989, 1991), Kim and Levin (1992), Kim and Tesanovic (1992, 1993). A similar result has been found for the one-band Hubbard model by Kulic and Zeyher (1994) and Mierzejewski *et al*. (1999).

[12] Resistivity of the form $\rho = \rho_0 + AT^n$ has been found in many studies, typically from $T_c$ to about 200K (Finnemore *et al*. 2001, Canfield *et al*. 2001, Chen *et al*. 2001, Pradhan *et al*. 2001, Schneider *et al*. 2001, Sologubenko *et al*. 2002, Eltsev *et al*. 2002, Kim *et al*. 2002, Putti *et al*. 2002). The exponent n differs

considerably among these studies, ranging from 1.9 to 3.0, but these exponents are all clearly smaller than the standard low-temperature behavior of Bloch-Gruneisen theory, where n = 5. [But there are other studies which have found conventional Bloch-Gruneisen behavior (Bharathi *et al.* 2002, Fisher *et al.* 2003, Masui and Tajima 2003). Nevertheless, Masui and Tajima (2003) found an improved fit by adding an n = 3 component, namely, by including an Einstein term to represent the prominent $E_{2g}$ phonon modes.] This variation of n between 2 and 3 for different samples has been found previously in exotic superconductors and also in some "reasonable candidate" materials such as VN, and this has a theoretical explanation in terms of a surprising enhancement of electron-electron scattering by anisotropic-scattering defects; see §4.8 of Brandow (2000) for discussion and references. However, in contrast to the established exotic superconductors the value of ρ at 300K is *not* several times larger than the corresponding value for lead (the prototype strong-coupling material); the ρ(300K) for $MgB_2$ is only around 10 μΩcm, definitely smaller than the lead value of 21 μΩcm. These features might be consistent with the general exotic phenomenology if one of the bands (σ or π) has a strong $T^2$ resistivity while the other band has conventional Bloch-Gruneisen behavior, with their conductivity contributions combining in a parallel manner as in Mazin *et al.* (2002). This resistive power-law behavior and its sample variation clearly need further study.

[13] The motivation for this remark is explained in §7 of Brandow (2000), where it is shown that this feature may facilitate the valence-fluctuation pairing mechanism. We expect that the partially occupied 3s orbitals of magnesium should carry a substantial Hubbard U parameter, in anlogy to the 6s orbitals of bismuth in BKBO. [The U for BKBO has been calculated by Vielsack and Weber (1996).]

[14] The exceptions are uranium heavy-fermion materials and $Sr_2RuO_4$, as explained in §1. See § 5.2 of Brandow (2000) for further discussion of this issue.

[15] After completion of Brandow (2002) we have found additional data showing clear

inner-gap features in cuprate tunneling data: Tao *et al*. (1991), Buschmann *et al*. (1992), Jeong *et al*. (1994), Ozyuzer *et al*. (2000). A particularly prominent example of inner-gap structure for Bi-2212 is shown in Mourachkine (1999), and is reproduced in Zhao (2001b).

[16] For simplicity we are not distinguishing between the surface field $H^*_{c1}$ of Ramirez (1996) and the true (bulk) $H_{c1}$.